\documentclass[twocolumn, 10pt]{IEEEtran}
\IEEEoverridecommandlockouts
\usepackage{graphicx}
\usepackage{float}
\usepackage{verbatim}
\usepackage{amsmath}
\usepackage{algpseudocode}
\graphicspath{ {image/} }
\usepackage{amsthm,amsmath,amssymb}
\usepackage{mathrsfs}
\usepackage{algorithm}
\usepackage{color}
\usepackage{verbatim}
\usepackage{xcolor}
\usepackage{cite}
\usepackage[separate-uncertainty = true,multi-part-units = repeat, load-configurations = abbreviations]{siunitx}
\usepackage{siunitx}
\usepackage{subfigure}
\ifCLASSINFOpdf
\else
\fi
\begin{document}
\title{Task-Oriented Semantics-Aware Communication for Wireless UAV Control and Command Transmission}

\author{Yujie Xu, Hui Zhou, and Yansha Deng% <-this % stops a space
%%%%%%%%%%%%%%%%%%%%%%%%%%%%%%%%%%%%%%%%%%%%%%%%%%%%%%%%thanks%%%%%%%%%%%%%%%%%%%%%%%%%%%%%%%%%%%%%%%%%%%%%%%%
\thanks{Yujie Xu, Hui Zhou and Yansha Deng are with the Department
of Engineering, King's College London, London, U.K.(e-mail: \{yujie.xu, hui.zhou,  yansha.deng\}@kcl.ac.uk)(Corresponding author: Yansha Deng).}

\thanks{This work was supported in part by the Engineering and Physical Research Council (EPSRC), U.K., under Grant EP/W004348/1. This work is also a contribution by  Project REASON, a UK Government funded project under the Future Open Networks Research Challenge (FONRC) sponsored by the Department of Science Innovation and Technology (DSIT).}}

\maketitle

\begin{abstract}
To guarantee the safety and smooth control of Unmanned Aerial Vehicle (UAV) operation, the new control and command (C\&C) data type imposes stringent quality of service (QoS) requirements on the cellular network. However, the existing bit-oriented communication framework is already approaching the Shannon capacity limit, which can hardly guarantee the ultra-reliable low latency communications (URLLC) service for C\&C transmission. To solve the problem, task-oriented semantics-aware (TOSA) communication has been proposed recently by jointly exploiting the context of data and its importance to the UAV control task. However, to the best of our knowledge, an explicit and systematic TOSA communication framework for emerging C\&C data type remains unknown. Therefore, in this paper, we propose a TOSA communication framework for C\&C transmission and define its value of information based on both the similarity and age of information (AoI) of C\&C signals. We also propose a deep reinforcement learning (DRL) algorithm to maximize the TOSA information. Last but not least, we present the simulation results to validate the effectiveness of our proposed TOSA communication framework.
\end{abstract}


% \fi
\IEEEpeerreviewmaketitle

\section{Introduction}
\IEEEPARstart{U}{NMANNED} Aerial Vehicle (UAV) has been regarded as an emerging technology to tackle a wide range of complicated tasks such as aerial imaging, food delivery, and traffic monitoring \cite{UAV_1, UAV_2, UAV_3, UAV_4}. To guarantee its safety operation, downlink control and command (C\&C) transmission imposes stringent quality of service (QoS) requirements for high reliability and low-latency. However, with the aim to achieve smooth control over complicated tasks, dramatically increasing C\&C data under high transmission frequency brings a heavy burden on the existing bit-oriented cellular network design \cite{UAV_uplink}.

Most existing work mainly focused on providing ultra-reliable low latency communications (URLLC) service for UAV communication \cite{URLLC_1, URLLC_2, URLLC_3}. In \cite{URLLC_1}, the average achievable data rate was derived for a UAV communication system under short packet transmission using 3-D channel model. In \cite{URLLC_2,URLLC_3}, the joint power and blocklength optimization and joint pilot and payload power optimization were performed for URLLC services in single antenna and massive-MIMO  UAV systems, respectively.  However, these works are developed based on traditional Shannon’s bit-oriented communication framework, which ignored the context and importance of each bit.

To fill the gap, task-oriented semantics-aware (TOSA) communication has been proposed as an emerging communication paradigm shift to semantic level and effectiveness level by jointly exploiting the context of data and its importance to the task \cite{task_aware}. In \cite{sc_1} and \cite{sc_2}, the authors proposed a robust semantic communication system for speech transmission. In \cite{ sc_3} deep learning enabled semantic communication system has been designed for textual data transmission.  However, existing works mainly focused on identifying the content of the traditional data type, including text and speech, as a semantic metric, where the data importance to the task has not been exploited. More importantly, the specific TOSA communication framework for emerging C\&C data type has never been proposed yet.

Motivated by this, in this paper, we develop a general TOSA communication framework for UAV C\&C transmission. The main contributions of this paper are threefold. First, to the best of our knowledge, it is the first paper that includes the semantics level and effectiveness level for emerging C\&C data type communication. Second, we define age of information (AoI) and similarity to quantify the semantic-level and effectiveness level performance, respectively. Third, we propose a general DRL algorithm to optimize the TOSA information of the downlink C\&C transmission based on the similarity and AoI of C\&C data.

The rest of this paper is organized as follows. Section II presents the system model and problem formulation. Section III introduces DRL-based TOSA communication framework for C\&C transmission. Section IV provides the simulation results. Finally, Section V concludes the paper.

\section{System Model And Problem Formulation}
\subsection{System Model}
As shown in Fig. 1, we assume that the ground control station(GCS) remains static all the time, and UAV user equipment(UAV-UE) flies in a circular horizontal disk with radius $\mathrm{R}$ and height $\mathrm{H}$, where the UAV-UE receives the C\&C transmission via the downlink. We mainly consider downlink C\&C transmission of this cellular-connected UAV network, where the base station(BS) serves as the GCS to periodically send the C\&C signal to the UAV-UE.

\begin{figure}[!h]
\vspace{-0.5cm}
\centering
\includegraphics[scale=0.2]{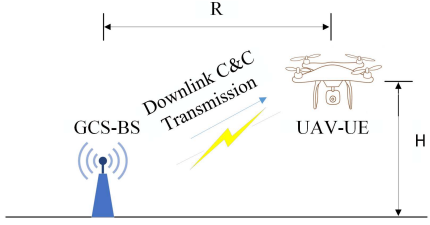}
\centering
\vspace{-0.2cm}
\caption{Diagram of transmission model between UAV-UE and GCS.}
\label{sys_mod}
\vspace{-0.5cm}
\end{figure}
Taking into account the potential line-of-sight (LoS) and non-line-of-sight (NLoS) for flying UAVs, we adopt free-space path loss and Rayleigh fading to model the path loss from the GCS to the UAV-UE as
\begin{equation} \label{h_mn_k}
\setlength\abovedisplayskip{1pt}
\setlength\belowdisplayskip{1pt}
h = \left\{
\begin{array}{lr}
\left(\frac{4\pi d f_{\mathrm{c}}}{c}\right)^{\alpha}\eta_{\mathrm{LoS}}\beta, \quad P_{\mathrm{LoS}}\\
\left(\frac{4\pi d f_{\mathrm{c}}}{c}\right)^{\alpha}\eta_{\mathrm{NLoS}}\beta, \quad P_{\mathrm{NLoS}} = 1 - P_{\mathrm{LoS}}\! \\
\end{array}
,
\right.
\end{equation}
where $d$ is the distance between UAV-UE and GCS, $f_c$ represents the downlink transmission frequency, $\eta_{\mathrm{LoS}}$ and $\eta_{\mathrm{NLoS}}$ are the path loss coefficients in LoS and NLoS cases, repectively. The $c$ represents the speed of light, $\alpha$ is path loss exponent, and $\beta$ denotes the Rayleigh small-scale fading following $\mathcal{CN}(0,1)$. In Eq. \eqref{h_mn_k}, $P_{\mathrm{LoS}}$ represents the LoS probability as
\begin{equation} \label{P_LoS_mn}
\setlength\abovedisplayskip{1pt}
\setlength\belowdisplayskip{1pt}
P_{\mathrm{LoS}} = \frac{1}{1+a\exp(-b(\theta - a))},
\end{equation}
where $\theta=\frac{180}{\pi}\arcsin{\frac{\mathrm{H}}{d}}$ is the elevation angle of the UAV-UE. The $\mathrm{H}$ represents the flight height of UAV-UE, $d$ is the distance between UAV-UE and GCS, and $\mathrm{a}$ and $\mathrm{b}$ are positive constants determined by the communication where $\mathrm{a}=11.95$ and $\mathrm{b}=0.14$ in dense urban environment. Based on Eq. \eqref{h_mn_k} and Eq. \eqref{P_LoS_mn}, the downlink channel from the GCS to the UAV-UE can be derived as
\begin{equation} \label{h_mn_k1}
\setlength\abovedisplayskip{1pt}
\setlength\belowdisplayskip{1pt}
h=(P_{\mathrm{LoS}}\eta_{\mathrm{LoS}}+P_{\mathrm{NLoS}}\eta_{\mathrm{NLoS}})\left(\frac{4 \pi d f_{\mathrm{c}}}{c}\right)^{\alpha}\beta.
\end{equation}

Therefore, we formulate the SNR as
\begin{equation}
\setlength\abovedisplayskip{1pt}
\setlength\belowdisplayskip{1pt}
\mathrm{SNR}=\frac{\mathrm{P} h}{\sigma^{2}},
\label{eq:SINR}
\end{equation}
where $\mathrm{P}$ represents the transmit power of BS,  $h$ is the channel gain, and $\sigma^2$ is Additive White Gaussain Noise (AWGN) power. It is noted that the UAV-UE can only successfully decode the C\&C message when the $\mathrm{SNR}$ of C\&C transmission is above the pre-defined threshold $\gamma_{th}$.

%\textcolor{red} {All five data have 32 bits to transmit, which means that one C\&C occupies 320 bits in total, 160 bits for the package header and 160 bits for data. BS accepts and processes one CC every time, and BS will immediately accept and process another control signal after one CC has been processed and transmitted. BS will receive one acknowledge character(ACK) if UAV-UE receives the control signal successfully.}

\subsection{Problem Formulation}
As discussed above, the BS periodically generates the C\&C signal and transmits to the UAV-UE. In this work, we tackle the problem of optimizing the scheduling decision defined by action parameter $A^t=\lambda^{t}$ for downlink C\&C transmission in each transmission time interval (TTI), where the $\lambda^{t}$ represents whether the BS decides to transmit or drop the C\&C signal in the $t^{\mathrm{th}}$ TTI. In order to select the action at the beginning of every TTI $t$, the BS accesses all prior historical observations $U^{t^{'}}$ in TTIs $ t^{'} =1,...t-1$ consisting of the following variables: the similarity $L^{t^{'}}$ between adjacent C\&C signals, and the age of information (AoI) $I^{t^{'}}$. We denote $O^{t}=\left\lbrace A^{t-1}, U^{t-1},A^{t-2}, U^{t-2},...,A^{1}, U^{1}\right\rbrace$ as the observed history of all such measurements and past actions. 
 
 In this paper, we aim at maximizing the long-term average reward $R^t$ related to the TOSA information for the downlink C\&C transmission in cellular-connected UAV network. The optimization relies on the selection of parameter in $A^t$ according to the current observation history $O^{t}$ with respect to the stochastic policy $\pi$. This optimization problem can be formulated as
	\begin{equation}
 \setlength\abovedisplayskip{1pt}
\setlength\belowdisplayskip{1pt}
(\text {P1})\!:\quad \mathop {\textbf {max}}\limits _{ \{\pi (A^{^{t}}|O^{^{t}})\}}~\sum _{k=t}^{\infty } \gamma ^{(k-t)} {\mathbb E}_{\pi }[R^t],
 \label{eq:problem}
	\end{equation}
	where $\gamma \in [0,1)$ is the discount rate for the performance in future TTIs. The decision process in GCS can be defined as MDP because the state $S^t$ including AoI, similarity and transmission result of similar packets is only related to its previous state and current action $A^t$ including transmission or drop. It is mentioned that the transmission result of similar packets can be achieved by similarity with the last packet and transmission result of similar packets in previous state. Since the GCS has no information on channel situation and do not know if each transmission will be successful or not, this introduces a Partially Observable Markov Decision Process (POMDP) problem, which is generally intractable. Approximate solutions will be discussed in Section \ref{DRL}.

\section{DRL-based TOSA Communication Framework for C\&C transmission}
\label{DRL}
%In order to solve equation (\ref{equ:sigma_reward}), in this section, we design one DRL algorithm to compare with the existing conventional downlink transmission methods. This DRL algorithm is capable of optimizing all the control signals in BS. In the following,...%%%%%%%%%%%%%%%%%%%%%%%

%DRL is one of the most capable methods to optimally transmit C\&Cs under a constantly changing communication environment because the deep neural network is one of the most impressive non-linear approximation functions[\ref{}]. The DRL algorithms have been widely deployed for dynamic optimization on wireless communication systems, e.g.,[\ref{}],[\ref{}]. However, there is no help to achieve an optimal solution using the direct application of existing DRL approaches.

In this section, we introduce the DRL-based approach to solve the problem in \eqref{eq:problem}. To maximize the TOSA information in downlink C\&C transmission, an agent is deployed at BS to choose the appropriate actions by exploring the environment progressively.  We define $s \in \mathcal{S}$, $a \in \mathcal{A}$, and $r \in \mathcal{R}$ as any state action, and reward from their corresponding sets, respectively. At the beginning of the $t$th TTI $\left(t\in\left\lbrace0,1,2,...\right\rbrace\right)$, the agent first observes the current state $S^{t}$ corresponding to a set of previous observations $O^{t}$ in order to select a specific action $A^{t}\in \mathcal{A}\left(S^{t}\right)$. The action $A^{t}$ is designed to be the scheduling decision $\lambda^t$.

We consider a basic state function in downlink C\&C transmission, where $S^{t}$ is a set of indices mapping to the current observed information $U^{t-1}=\left[I^{t-1}, L^{t-1}\right]$. With the knowledge of the state $S^{t}$, the agent chooses an action $A^{t}$ from the set $\mathcal{A}$, which is a set of indices mapped to the set of available scheduling decision $\mathcal{F}=\{0,1\}$. Once an action $A^{t}$ is performed, the agent will receive a scalar reward $R^{t+1}$, and observe a new state $S^{t+1}$. The reward $R^{t+1}$ indicates to what extent the executed action $A^{t}$ can achieve the optimization goal, which is determined by the new observed state $S^{t+1}$. As the optimization goal is to maximize the TOSA information, we define the reward $R^{t+1}$ as a function related to the observed similarity between adjacent C\&C signals $L^t$ and $L^{t-1}$ and the AoI $I^t$ and $I^{t-1}$, which is defined as
\begin{equation}
\setlength\abovedisplayskip{1pt}
\setlength\belowdisplayskip{1pt}
\begin{aligned}
\begin{split}
R^{t+1}= \left \{
\begin{array}{ll}
    f(L^t) g(I^t)                   & \text{Successful transmission}\\
    0                           & \text{Failure transmission}.
\end{array}
,
\right.
\end{split}
\end{aligned}
\label{equ:reward}
\end{equation}
In Eq. \ref{equ:reward}, the similarity quantifies the importance of C\&C related to the task, and AoI quantifies the freshness of the C\&C signal, both of which will be discussed below.

To quantify the difference between consecutive C\&C signals, we define the similarity as 
\begin{equation}
\setlength\abovedisplayskip{1pt}
\setlength\belowdisplayskip{1pt}
L^t=\sum_i (\mu_i \frac{M^t_i-M^{t-1}_i}{R_i}),
\label{equ:sim}
\end{equation}
where $i \in \lbrace \text{ROW, PITCH, YAW, THRUST} \rbrace$, $\mu_i$ is the weight for UAV C\&C parameter $i$, and $M^t_i$ represents the value of parameter $i$ at the $t$th TTI, which can be obtained from C\&C packets at $t$th TTI directly and some important information of $M^t_i$ is shown in Table \ref{Mi}. Moreover, in Eq. \eqref{equ:sim}, $\frac{M^t_i-M^{t-1}_i}{R_i}$ describes the gap of UAV C\&C parameter $i$ between two C\&C signals in the range of [0,1]. To normalize the similarity value, we utilize the sigmoid function as
\begin{equation}
\setlength\abovedisplayskip{1pt}
\setlength\belowdisplayskip{1pt}
 f(L^t)=\frac{2}{1+e^{-\kappa (L^t-\zeta)}} - 1,
\label{equ:s}
\end{equation}
where $\kappa$ is the parameter controlling the gain, and $\zeta$ controls the cutoff.
\begin{table}[h!]
\centering
\vspace{-0.5cm}
\caption{Information of $M^t_i$ and $R_i$}
\vspace{-0.3cm}
\begin{tabular}{|c | c| c|c|} 
 \hline
 $i$ & $\mathbf{Length}$ & $M^t_i~\mathbf{Range}$ & $R_i$\\
 \hline
 ROW & 4 bytes & $-35^{\circ}$ to $35^{\circ}$& 70\\
  \hline
 PITCH & 4 bytes & $-35^{\circ}$ to $35^{\circ}$ & 70\\
  \hline
  YAW & 4 bytes & $-150^{\circ}/s$ to $150 ^{\circ}/s$ & 300\\
  \hline
  THRUST & 4 bytes & $-5 m/s$ to $5 m/s$ & 10\\
 \hline
\end{tabular}
\vspace{-0.45cm}
\label{Mi}
\end{table}
The AoI represents the freshness of the transmitted C\&C signal. When the C\&C signal is successfully received, the value of AoI is derived as
\begin{equation}
\setlength\abovedisplayskip{1pt}
\setlength\belowdisplayskip{1pt}
I^t=t_{\rm rcv}-t_{\rm gen},
\label{equ:ins_AoI}
\end{equation}
where $t_{\rm rcv}$ is the time that the packet is received by the UAV-UE and confirmed by ACK, and  $t_{\rm gen}$ is the generated time of C\&C signal at the GCS. Eq.\eqref{equ:ins_AoI} manifests that $I^t$ will increase with the lifetime of the C\&C signals, which can be described as the freshness of the data. It is noted that Eq. \eqref{equ:ins_AoI} also represents the transmission time of the newly generated C\&C packet, which can be derived as 
\begin{equation}
\setlength\abovedisplayskip{1pt}
\setlength\belowdisplayskip{1pt}
    I^t=\frac{\mathrm{N_{cc}}}{B \log(\mathrm{SNR}+1)},
    \label{eq:t_trans}
\end{equation}
where $\mathrm{N_{cc}}$ is the size of C\&C signal, and $B$ is bandwidth.

Without loss of generality, we assume the transmission of each C\&C signal completes within each TTI. Hence, we normalize the $I^t$ as 
\begin{equation}
\setlength\abovedisplayskip{1pt}
\setlength\belowdisplayskip{1pt}
    g(I^t)=1-\frac{I^{t}}{\Delta T},
    \label{equ:VoI}
\end{equation}
where $\Delta T$ is the duration of each TTI.

	We consider $\epsilon$-greedy approach to balance exploitation and exploration in the Actor of the Agent, where $\epsilon$ is a positive real number and 
	$\epsilon\leq 1$. In each TTI $t$, the agent randomly generates a probability $p^{t}_{\epsilon}$ to compare with $\epsilon$. Then, with the probability $\epsilon$, the algorithm randomly chooses an action from the remaining feasible actions to improve the estimate of the non-greedy action’s value. With the probability $1-\epsilon$, the exploitation is obtained by performing forward propagation of Q-function $Q(s, a; \theta)$ with respect to the observed state $S^{t}$. The weights matrix $\theta$ is updated online along each training episode by using double deep Q-learning (DDQN), which to some extent reduce the substantial overestimations of value function. Accordingly, learning takes place over multiple training episodes, with each episode of duration $N_{\mathrm{TTI}}$ TTI periods. In each TTI, the parameter $\theta$ of the Q-function approximator $Q(s, a; \theta)$ is updated using RMSProp optimizer as
	\begin{equation} 
 \setlength\abovedisplayskip{1pt}
\setlength\belowdisplayskip{1pt}
	{\boldsymbol \theta }^{t+1} = {\boldsymbol \theta }^{t} - \lambda _{\text {RMS}} \nabla L({\boldsymbol \theta }^{t}),
	\label{eq:optimizer}
	\end{equation}
	where $\lambda _{\text {RMS}}$ is RMSProp learning rate, $\nabla L({\boldsymbol \theta })$ is the gradient of the loss function $L({\boldsymbol \theta })$ used to train the Q-function approximator. This is given as
	\begin{equation}
 \setlength\abovedisplayskip{1pt}
\setlength\belowdisplayskip{1pt}
\begin{small}
	\begin{aligned}
	\nabla L(\boldsymbol{\theta} ^{t})=&{\mathbb E}_{S^{i},A^{i},R^{i+1},S^{i+1}} \big [\!\big (\!R^{i+1}\!+\! \gamma \mathop {\text {max}}\limits _{a}Q(S^{i+1}, a; \bar {\boldsymbol {\theta }}^{t}) \\&\qquad \,\!-\,Q(S^{i}, A^{i}; \boldsymbol {\theta }^{t}) \big) \nabla _{\boldsymbol {\theta }} Q(S^{i}, A^{i}; \boldsymbol {\theta }^{t})\big],
	\label{eq:DQN_update}
	\end{aligned}
\end{small}
	\end{equation}
	where the expectation is taken with respect to a so-called minibatch, which are randomly selected previous samples $(S^{i},A^{i},S^{i+1},R^{i+1})$ for some $i \in \lbrace t-M_r,...,t \rbrace$, with $M_r$ being the replay memory. When $t-M_r$ is negative, this is interpreted as including samples from the previous episode. The use of minibatch, instead of a single sample, to update the value function $Q(s, a; \theta)$ improves the convergent reliability of value function. Furthermore, following DDQN, in \eqref{eq:DQN_update}, $\bar {\boldsymbol {\theta }}^{t}$ is a so-called target Q-network that is used to estimate the future value of the Q-function in the update rule. This parameter is periodically copied from the current value $\boldsymbol {\theta }^{t}$ and kept fixed for a number of episodes. In the following, the implementation of DRL-based TOSA Communication Framework for C\&C transmission is shown in \textbf{Algorithm \ref{Centralized_DQN}}. This algorithm has the time complexity of $O(nTn_{in}n_ln_h^2)$, where $n$ is episode, $T$ is the total TTI in each episode, $n_{in}$ is the input size, $n_l$ is the number of layers in GRU and $n_h$ is the hidden size.

\begin{algorithm}[h]
\caption{DRL-based TOSA Communication Framework for C\&C transmission}
\label{Centralized_DQN}
\hspace*{0.02in}{\bf Input:}
The set of available action $\mathcal{F}$.
\begin{algorithmic}[1]
\State Algorithm hyperparameters: learning rate $\lambda_{\text {RMS}} \in (0,1]$, discount rate $\gamma \in (0,1]$, $\epsilon$-greedy rate $\epsilon \in (0,1]$, target network update frequency $K$\;
\State Initialization of replay memory $M$ to capacity $C$, the primary Q-network $\boldsymbol{\theta}$, and the target Q-network $\boldsymbol{\overline{\theta}}$\;
\For {$t=1,...,T$}
    \State Update the traffic\;
    \If {$p_{\epsilon}^{t}<\epsilon$}
        \State select a random action $A^{t}$ from $\mathcal{A}$\;
    \Else
        \State select $A^{t}=\mathrm{argmax}Q(S^{t},a,\boldsymbol{\theta})$\;
    \EndIf
    \State The BS executes the decision, transmits $t$th C\&C signal or drops $t$th C\&C signal\;
    \State The central server observes $S^{t+1}$, and calculate the related $R^{t+1}$ using Eq.~\ref{equ:reward}\;
    \State Store transition $(S^{t}, A^{t}, R^{t+1}, S^{t+1})$ in replay memory $M$\;
    \State Sample random minibatch of transitions $(S^{t}, A^{t}, R^{t+1}, S^{t+1})$ from replay memory $M$\;
    \State Perform a gradient descent for $Q(s,a,\boldsymbol{\theta})$ using (\ref{eq:DQN_update})\;
    \State Every $K$ steps update target Q-network $\overline{\boldsymbol{\theta}}=\boldsymbol{\theta}$.
\EndFor
\end{algorithmic}
\label{alg:DQN}
\end{algorithm}

\section{Simulation Results}
In this section, we present the simulation results of DRL-based TOSA communication for C\&C transmission. In simulation, we set carrier frequency $f_c$ as 5GHz, SNR threshold $\gamma_{th}$ as 5.5 dB, the noise $\sigma^2$ as -104 dBm and the transmit power $P$ as 18 dBm. In the centralized DDQN algorithm, we set the greedy rate $\epsilon$ as 1. The batch size is 32, the learning rate $\lambda_{\text {RMS}}$ is $10^{-5}$ and the discount rate $\gamma$ is 0.1. We obtain the average testing performance over 20 groups of C\&C data sets at the GCS. In C\&C data sets $k=1,2,...,20$, every data set has $\mathrm{N}$ different C\&C packets and every different packet is followed by $k-1$ same packets, which means that there are $k$ consecutive same packets and the number of total packets in data set $k$ is $k\mathrm{N}$. It is mentioned that $k$ is one parameter that describes the repeat times in data set instead of retransmission index. In our simulation, we name multiple consecutive repetitive C\&C packets as one effective C\&C packet because the consecutive repetitive C\&C packets transmit the same meaning of data. Take one example of $\mathrm{N}=3$ and $k=4$, then the data set can be  [$30^{\circ}/s$, $30^{\circ}/s$, $30^{\circ}/s$, $30^{\circ}/s$, $0^{\circ}/s$, $0^{\circ}/s$, $0^{\circ}/s$, $0^{\circ}/s$, $30^{\circ}/s$, $30^{\circ}/s$, $30^{\circ}/s$, $30^{\circ}/s$] for $\mathrm{YAW}$ with $\mathrm{THRUST}=3m/s, \mathrm{ROW}=0, \mathrm{PITCH}=0$, where the effective C\&C packets are 3. In this simulation, we deploy one GCS with TOSA communication and one UAV-UE.
\begin{comment}
\begin{table}[h!]
\centering
\caption{One Example of $\mathrm{N}=3$ and $k=4$}
\begin{tabular}{|c | c| c|c|} 
 \hline
 $\mathbf{Sequence}$ & $\mathbf{Thrust}$ & $\mathbf{Yaw}$ & $\mathbf{Roll \& Pitch }$\\
 \hline
 $1$ & $3m/s$ & $0$ & $0$\\
\hline
$2$ & $3m/s$ & $0$ & $0$\\
\hline
$3$ & $3m/s$ & $0$ & $0$\\
\hline
$4$ & $3m/s$ & $0$ & $0$\\
\hline
$5$ & $3m/s$ & $30^{\circ}/s$ & $0$\\
\hline
$6$ & $3m/s$ & $30^{\circ}/s$ & $0$\\
\hline
$7$ & $3m/s$ & $30^{\circ}/s$ & $0$\\
\hline
$8$ & $3m/s$ & $30^{\circ}/s$ & $0$\\
\hline
$9$ & $-3m/s$ & $0^{\circ}/s$ & $0$\\
\hline
$10$ & $-3m/s$ & $0^{\circ}/s$ & $0$\\
\hline
$11$ & $-3m/s$ & $0^{\circ}/s$ & $0$\\
\hline
$12$ & $-3m/s$ & $0^{\circ}/s$ & $0$\\
\hline
\end{tabular}
\label{data_set}
\end{table}
\end{comment}
\begin{comment}
\begin{table}[h!]
\centering
\caption{Parameters}
\begin{tabular}{|c | c| c|c|} 
 \hline
 $\mathbf{Parameter}$ & $\mathbf{Value}$ & $\mathbf{Parameter}$ & $\mathbf{Value}$\\
 \hline
 $\eta_{\mathrm{LoS}}$/$\eta_{\mathrm{NLoS}}$& $10^{-0.1}$/$10^{-2}$ & $\mathrm{N}$ & 10000\\
 \hline
$P$ & 18dBm & $\gamma$ & 0.1\\ 
\hline
$\alpha$ & -2&$\lambda_{\text {RMS}}$ & $10^{-5}$\\
\hline
$f_c$ & 5GHz & $\sigma^2$ & -104dBm\\
\hline
$\gamma_{th}$ & 5.5dB& $B$ & 20Mbps\\
\hline
\end{tabular}
\label{para}
\end{table}
\end{comment}
\begin{figure}[H]
\vspace{-0.3cm}
\centering
\includegraphics[scale=0.65]{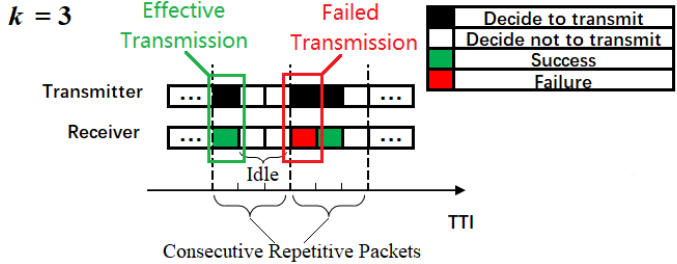}
\centering
\vspace{-0.6cm}
\caption{Agent decisions with repetitive times $k=3$.}
\label{fig:timeline}
\vspace{-0.3cm}
\end{figure}
Fig. \ref{fig:timeline} presents the scheduling decisions in our proposed TOSA communication with repetitive times $k=3$. We can observe that the agent learns to transmit the new different C\&C signal once it has arrived. If the C\&C signal has been successfully received by the receiver, which means an effective transmission occurs, the agent will decide to not transmit the consecutive similar C\&C signals. Otherwise, if the agent experiences a failed transmission, the agent will continue to schedule the downlink C\&C transmission of the next similar signal until one effective transmission occurs or a new different C\&C signal arrives.
\begin{figure}[H]
\vspace{-0.3cm}
\centering
\subfigure[%Number of C\&C transmission times of both bit-oriented communication and TOSA communication with different repeat times $k$ for each effective C\&C signal.
]{
\includegraphics[scale=0.343]{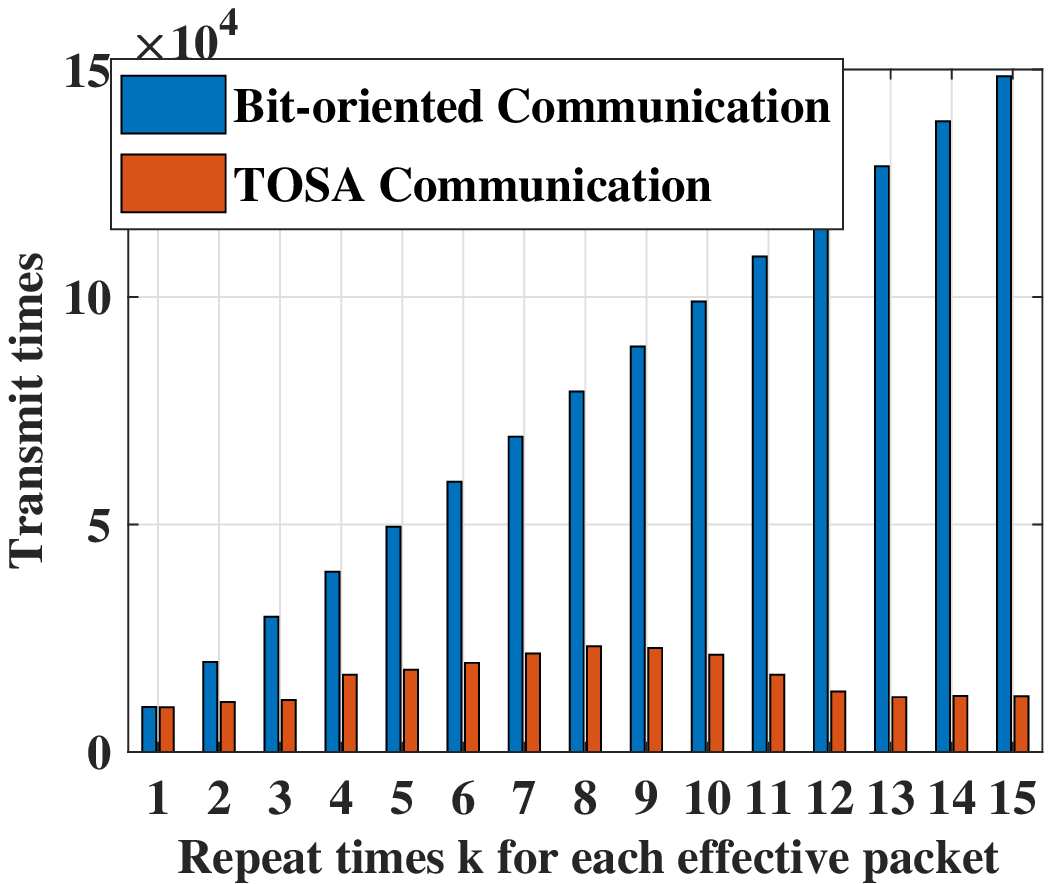}
\label{num_of_attempts}
}
\subfigure[%Effective transmission rate versus various repeat times $k$ for each effective C\&C signal.
]{
\includegraphics[scale=0.343]{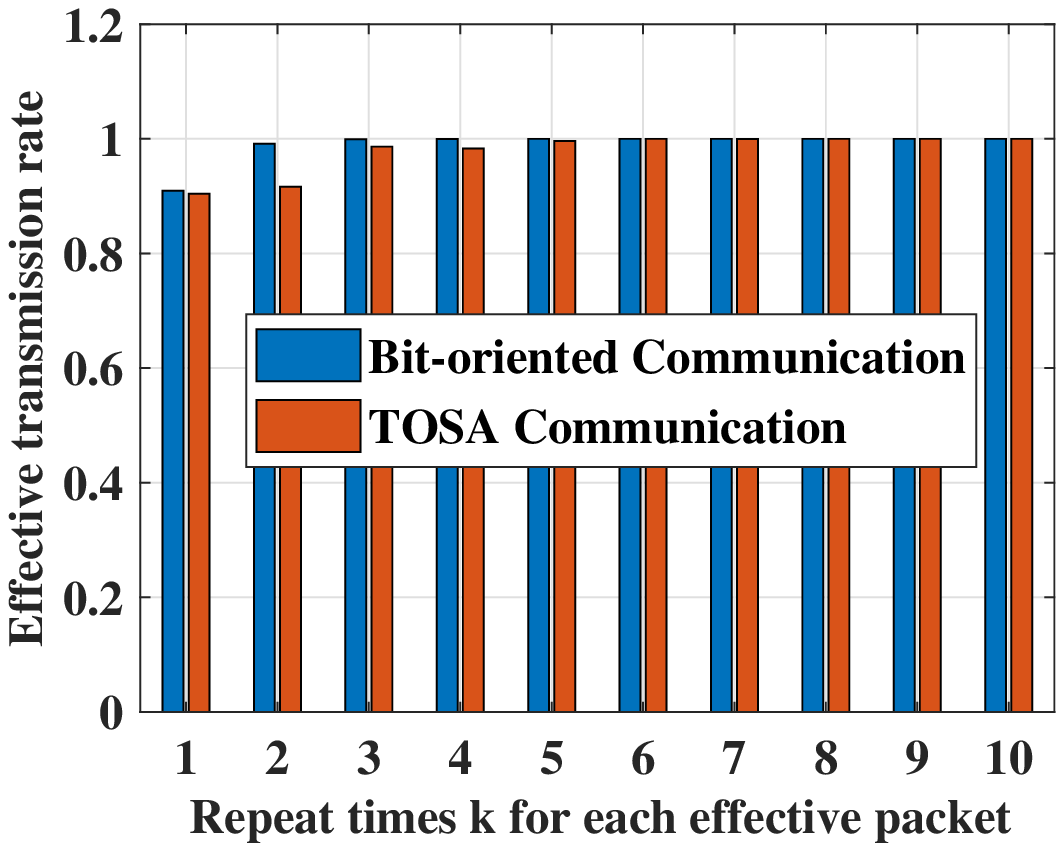}
\label{suc_rate}
}
\vspace{-0.3cm}
\centering
\caption{Simulation results with different repeat times $k$: (a) Number of C\&C transmission times of both bit-oriented communication and TOSA communication with different repeat times $k$ for each effective C\&C signal. (b) Effective transmission rate versus various repeat times $k$ for each effective C\&C signal.}
\vspace{-0.3cm}
\end{figure}
Fig. \ref{num_of_attempts} compares the number of transmission times between the conventional bit-oriented communication and proposed TOSA communication. With the increase of the same packet repetitive times $k$, the actual number of transmission times remains stable in our TOSA framework, which is slightly larger than the number of effective packets due to transmission failure. Compared to the conventional bit-oriented communication, the BS learns to only transmit the important C\&C signal and drop the other less important C\&C signal, which leads to much lower transmission times.

Fig. \ref{suc_rate} plots the effective transmission rate, which is defined as the effective transmission times divided by the total effective packets $\mathrm{N}$, versus different repetitive times $k$ for each effective C\&C signal. We can observe that our proposed TOSA communication achieves almost the same effective transmission rate as traditional bit-oriented communication which transmits all C\&C signal packets. It is noted that the negligible performance gain of traditional bit-oriented comes from incredible resource consumption to transmit all the generated C\&C signals, which verifies that our proposed TOSA communication can guarantee the effectiveness of the  C\&C task execution with much lower resource consumption.

\section{Conclusion}
In this paper, we designed a general TOSA communication framework for UAV C\&C downlink transmission, where both semantic level and effectiveness level are incorporated. We defined the similarity and AoI of C\&C signals to quantify the effectiveness level and semantic level performances, respectively. We further proposed a DRL algorithm to maximize the TOSA information via optimizing the transmit or drop decisions at the transmitter. Our numerical results shed light on that our proposed TOSA framework can guarantee the C\&C task execution with much fewer communication resources.

% or



\ifCLASSOPTIONcaptionsoff
  \newpage
\fi

\end{document}